\documentstyle[epsf]{mn}

\newif\ifAMStwofonts

\def\k{\mbox{\boldmath$k$}}
\def\x{\mbox{\boldmath$x$}}
\def\y{\mbox{\boldmath$y$}}

\ifoldfss
  \ifCUPmtlplainloaded \else
    \NewTextAlphabet{textbfit} {cmbxti10} {}
    \NewTextAlphabet{textbfss} {cmssbx10} {}
    \NewMathAlphabet{mathbfit} {cmbxti10} {} 
    \NewMathAlphabet{mathbfss} {cmssbx10} {} 
  \fi
  \ifAMStwofonts
    \ifCUPmtlplainloaded \else
      \NewSymbolFont{upmath} {eurm10}
      \NewSymbolFont{AMSa} {msam10}
      \NewMathSymbol{\upi}     {0}{upmath}{19}
      \NewMathSymbol{\umu}     {0}{upmath}{16}
      \NewMathSymbol{\upartial}{0}{upmath}{40}
      \NewMathSymbol{\leqslant}{3}{AMSa}{36}
      \NewMathSymbol{\geqslant}{3}{AMSa}{3E}

      \let\geq=\geqslant 
    \fi
  \fi
\fi 

\ifnfssone
  \newmathalphabet{\mathit}
  \addtoversion{normal}{\mathit}{cmr}{m}{it}
  \addtoversion{bold}{\mathit}{cmr}{bx}{it}
  \newmathalphabet{\mathbfit} 
  \addtoversion{normal}{\mathbfit}{cmr}{bx}{it}
  \addtoversion{bold}{\mathbfit}{cmr}{bx}{it}
  \newmathalphabet{\mathbfss} 
  \addtoversion{normal}{\mathbfss}{cmss}{bx}{n}
  \addtoversion{bold}{\mathbfss}{cmss}{bx}{n}
  \ifAMStwofonts
    \ifCUPmtlplainloaded \else
      \UseAMStwoboldmath
      \makeatletter
      \new@mathgroup\upmath@group
      \define@mathgroup\mv@normal\upmath@group{eur}{m}{n}
      \define@mathgroup\mv@bold\upmath@group{eur}{b}{n}
      \edef\UPM{\hexnumber\upmath@group}
      \new@mathgroup\amsa@group
      \define@mathgroup\mv@normal\amsa@group{msa}{m}{n}
      \define@mathgroup\mv@bold\amsa@group{msa}{m}{n}
      \edef\AMSa{\hexnumber\amsa@group}
      \makeatother
      \mathchardef\upi="0\UPM19
      \mathchardef\umu="0\UPM16
      \mathchardef\upartial="0\UPM40
      \mathchardef\leqslant="3\AMSa36
      \mathchardef\geqslant="3\AMSa3E

      \let\geq=\geqslant 
    \fi
  \fi
\fi 

\ifnfsstwo
  \DeclareMathAlphabet{\mathbfit}{OT1}{cmr}{bx}{it}
  \SetMathAlphabet\mathbfit{bold}{OT1}{cmr}{bx}{it}
  \DeclareMathAlphabet{\mathbfss}{OT1}{cmss}{bx}{n}
  \SetMathAlphabet\mathbfss{bold}{OT1}{cmss}{bx}{n}
  \ifAMStwofonts
    \ifCUPmtlplainloaded \else
      \DeclareSymbolFont{UPM}{U}{eur}{m}{n}
      \SetSymbolFont{UPM}{bold}{U}{eur}{b}{n}
      \DeclareSymbolFont{AMSa}{U}{msa}{m}{n}
      \DeclareMathSymbol{\upi}{0}{UPM}{"19}
      \DeclareMathSymbol{\umu}{0}{UPM}{"16}
      \DeclareMathSymbol{\upartial}{0}{UPM}{"40}
      \DeclareMathSymbol{\leqslant}{3}{AMSa}{"36}
      \DeclareMathSymbol{\geqslant}{3}{AMSa}{"3E}

      \let\geq=\geqslant 
    \fi
  \fi
\fi 

\ifCUPmtlplainloaded \else
  \ifAMStwofonts \else 
    \def\upi{\pi}
    \def\umu{\mu}
    \def\upartial{\partial}
  \fi
\fi

\title[]
  {Constraints on a non-Gaussian ($\chi_m^2$) CDM model}
\author[]
  {K.Koyama,$^1$ J.Soda, $^2$ A.Taruya $^3$ \\
  $^1$ Graduate School of Human and Environment Studies, Kyoto University, 
       Kyoto  606-8501, Japan \\
  $^2$ Department of Fundamental Sciences, FIHS, Kyoto University,
       Kyoto, 606-8501, Japan \\
  $^3$ Research Center for the Early Universe (RESCEU)
       School of Science, University of Tokyo, 
       Bunkyo-ku, Tokyo 113-0033, Japan}

\date{1999 Nov 1.}
\pagerange{\pageref{firstpage}--\pageref{lastpage}}
\pubyear{}

\def\LaTeX{L\kern-.36em\raise.3ex\hbox{a}\kern-.15em
    T\kern-.1667em\lower.7ex\hbox{E}\kern-.125emX}

\begin{document}

\label{firstpage}

\maketitle

\begin{abstract}
We consider constraints on the structure formation model based on  
non-Gaussian fluctuations generated during inflation, which have 
$\chi_m^2$ distributions. Using three data sets, 
the abundance of the clusters at $z=0$, moderate $z$ and the correlation 
length, we show that constraints on the non-Gaussianity and the amplitude of 
fluctuations and the density parameter can be obtained. We obtain
an upper bound for $\Omega_m$ and a lower bound for the non-Gaussianity
and the amplitude of the fluctuations.
Using the abundance of clusters at $z \sim 0.6$,
for the spectrum parameterized by cold dark matter (CDM) shape 
parameter $\Gamma=0.23$, 
we obtain an upper bound for the density parameter $\Omega_m \sim 0.5$ and 
lower bounds for the amplitude $\sigma_8 \sim 0.7$ and for 
the non-Gaussianity of fluctuations $G \sim 2$ $(m \sim 200)$, 
where $G=1$ for Gaussian.

\end{abstract}


\section{Introduction}
It is generally accepted that the primeval density 
fluctuations have grown gravitationally to form the present structure 
of the Universe. Many models based on Gaussian adiabatic fluctuations 
generated during inflation have been discussed. However, the standard CDM 
model, i.e. the structure formation model in a spatially flat 
universe filled with cold dark matter which has Gaussian adiabatic 
fluctuations, suffers from several problems.
One of the most severe problems is the lack of the cluster abundance
at moderate $z$. This is caused by the fact that in a flat universe
filled with the matter, fluctuations grow rapidly. To overcome this problem, 
a low-density universe model, with or without cosmological constant 
has been investigated extensively. 

Recently an alternative was proposed,
a model based on non-Gaussian isocurvature fluctuations 
produced by massive scalar fields frozen during inflation,
which have $\chi_m^2$ distributions (Peebles 1997a,1999a,b). 
The most notable point of this model is the early formation of objects
because of its non-Gaussianity. 
If the probability distribution function (PDF) 
has a broad tail compared with a Gaussian PDF, the evolution of clusters 
becomes slow. 
In this model there are three important parameters; the 
amplitude and the non-Gaussianity of the fluctuations and the background 
density parameter. It is important
to explore methods of constraining these parameters.

It is recognized that the non-Gaussianity affects the correlation 
length of clusters. Robinson, Gawiser $\&$ Silk (1998) suggest 
that by combining the
observations of the abundance of clusters at $z=0$ and the correlation
length, constraints on the amplitude and the
non-Gaussianity of the fluctuations are obtained simultaneously
for a fixed density parameter $\Omega_m$.
As mentioned above, because the density parameter and the non-gaussanity 
affect the evolution of abundance, the observations of the abundance of
clusters at moderate $z$ give another strong constraint on the model. 
Much work on the effect of changing $\Omega_m$ on the evolution
have been carried out assuming Gaussian initial conditions
(Frenk et al. 1990; Eke et al. 1996; 1998; Viana $\&$ Liddle 1996,1998; 
Fan et al. 1997; Bahcall $\&$ Fan 1998;  Henry 1997; Carlberg et al 1997). 
Little work has been carried out, however, on the effect of the 
non-Gaussianity, except for a topological defect model 
(Chui $\&$ Ostricker 1998).

In this paper, we extend Press-Schechter (PS) theory 
to non-Gaussian initial conditions and make clear how the non-Gaussianity
affects the evolution of abundance and the bias. Then we show that 
combining the observations of the abundance of clusters
at moderate $z$ with the method proposed by Robinson.et.al, we can
obtain a constraint on background cosmology and fluctuations 
simultaneously: 
an upper bound for $\Omega_m$ and a lower bound for the non-Gaussianity
and the amplitude of the fluctuations. 

\section{Mass function and bias: for Non-Gaussian initial conditions}

We extend PS theory (Press $\&$ Schechter 1974; Bond et al.
1991) to non-Gaussian initial conditions and 
calculate the mass function and the bias of clusters. 

In PS theory, the
region of the scale $R$ with overdensity greater than a critical amount 
$\delta_c$ is collapsed to form a bound object of mass $M=(4 \pi/3)
\bar{\rho} R^3$, where $\bar{\rho}$ is the mean density of the universe. 
The threshold is assumed to be given by the spherical 
collapse model. For Gaussian initial conditions, 
it has been shown to be a good approximation (Lacey $\&$ Cole 1993). 
We assume that this approximation is good for non-Gaussian initial conditions.
The probability of a region with mass $M$ collapsing at
redshift $z_1$ is given by
\begin{equation}
F(\delta_1, M)=\int^{\infty}_{\delta_1} P(\delta) d \delta
,\quad \delta_1= \frac{\delta_c}{D(z_1)},
\end{equation}
where $P(\delta)$ is the PDF in question and $D(z)$ is a linear growth factor.
According to Jedamzick, the above quantity can be written as
\begin{equation}
F(\delta_1,M)=\frac{1}{\bar{\rho}}\int^{\infty}_{M} M'  n(M',z_1) P(M |M') dM',
\end{equation}
where  $n(M,z_1)$ is the mass function at $z=z_1$ and 
$P(M | M')$ denotes the probability of finding a region with mass $M$ 
overdense by $\delta_1$ or more, provided it is included in an isolated 
overdense region with mass $M'$ (Jedamzick 1995).
We assume this quantity is given by
\begin{equation}
P(M | M')= P(\delta_M \geq \delta_1 | \delta_{M'}=\delta_1)
=\int^{\infty}_{\delta_1} P(\delta | \delta_1) d \delta \equiv 1/f,
\end{equation}
where $P(\delta | \delta_1)$ is a conditional probability
function. Hence, the mass function $n(M,z_1)$ is given by
\begin{equation}
n(M,z_1)=- \frac{f \bar{\rho}}{M} \frac{d F(\delta_1, M)}{d M}.
\end{equation}

The bias of the bound objects can be calculated using the 
formalism developed by Mo and White (Mo $\&$ White 1996). 
Consider the collapsed region of scale $R_1$ with over-density
$\delta_1$ and the uncollapsed region of $R_0$ with $\delta_0$. 
If at present $\delta_1$ = $\delta_c /D(z_1)$, then the halo is identified 
as being formed at $z_1$. The number of objects of $M_1$ identified at
$z_1$ in a spherical region with a comoving radius $R_0$ is 
\begin{equation}
N(1|0)dM_1=-f \frac{M_0}{M_1} \frac{d}{d M_1} \int^{\infty}_{\delta_1}
d \delta P(\delta | \delta_0).
\end{equation}
Thus, the average overdensity of objects in the sphere of radius $R_0$ 
relative to the global mean halo abundance becomes
\begin{equation}
\delta^{L}_{cl} (1|0)=\frac{N(1|0)}{n(M,z) V_0} -1
\end{equation}
in Lagrangian space (Mo $\&$ White 1996). 

If the region $0$ is sufficiently large compared with region $1$
( $R_0 \gg R_1$, $\langle \delta_1^n \rangle \gg \langle \delta_0^n
\rangle $), regardless of the statistics, the relation
\begin{equation}
\langle \delta_1^p \delta_0^{n-p} \rangle = \langle \delta_0^n \rangle
\end{equation}
is satisfied (Appendix A) and the conditional probability is given by
\begin{equation}
P(\delta_1 |\delta_0)=P(\delta_1-\delta_0).
\end{equation}
(Appendix B). If this is the case, 
we obtain the normalization of the mass function
\begin{equation}
f=1/(\int^{\infty}_{0} P(\delta)d \delta),
\end{equation}
which agrees with the one introduced to ensure that
all the mass in the universe is accounted for 
(Lucchin and Matarrese 1989, Chui and Ostriker 1998). Since
\begin{equation}
N(1|0)dM=-f \frac{M_0}{M_1} \frac{d}{d M_1} \int^{\infty}_{\delta_1-\delta_0}
d \delta P(\delta),
\end{equation}
the overdensity of the bound objects is given by
\begin{equation}
 \delta^{L}_{cl} =\frac{n(1|0)-n(M)}{n(M)}, \quad\quad n(1|0)
 =n(M)|_{\delta_1 \to \delta_1 -\delta_0}.
\end{equation}
This reproduces the peak-background split formalism
(Kaiser 1984; Bardeen et al. 1986; Robinson et al. 1998).

Furthermore, we assume that reduced moments 
\begin{equation}
\lambda_N=\frac{\langle \delta^N \rangle}{\langle \delta^2 \rangle^{N/2}}
\end{equation}
do not depend on mass $M$ on the relevant scales.

In the following discussion, 
we make these assumptions and use above formalisms.
Note that the accuracy required to use Press-Schechter theory for non-Gaussian
initial conditions is currently uncertain. One way to confirm its validity 
is to perform N-body simulations. Recently Baker and Robinson have carried 
out N-body simulations for several non-Gaussian models to show mass function 
of PS formalism agrees well with that observed in
simulations (Robinson \& Baker1999). 
It will be necessary to perform further detailed tests of the accuracy of 
this formalisms, especially for the bias model.

\section{The abundance and the bias of the bound objects.}

We use the formalisms developed in the previous section and investigate the
effect of non-Gaussianity on the evolution of abundance and the bias 
of the bound objects. 
First we consider the evolution of the number density.
We define the rate of change with $z$ of the number density by
\begin{equation}
\triangle n(M,z) =-\frac{1}{n(M,z)} \frac{d n(M,z)}{d z}.
\end{equation}
This can be written as
\begin{equation}
\triangle n(M,z) = \left[-\frac{d \log \nu P(\nu)}{d \nu} \right]_{\nu(M,z)}
\left(-\frac{d \log D(z)}{dz} \right) \nu(M,z) ,
\end{equation}
where
\begin{equation}
\nu(M,z)=\frac{\delta_c}{\sigma(M) D(z)}, \quad \sigma(M)=\sigma_8
\left(\frac{M}{M_{\ast}} \right)^{-(3+n)/6},
\end{equation}
where $M_{\ast}$ is the mass contained in the spherical region of $8 h^{-1}
$ Mpc and $n$ is the spectrum index.
This indicates that for a fixed PDF, if fluctuations decrease slow with $z$
(as in low $\Omega_m$ universe), the number 
density also decreases slowly with $z$. We fix the background
cosmology and compare the evolution of the number density for different PDFs.
The difference of this quantity between a Gaussian PDF 
and the PDF in question can be written as
\begin{equation}
\triangle n_{G}-\triangle n= \left[\frac{d \log g(\nu)}{d \nu} 
\right]_{\nu(M)} \left(-\frac{d \log D(z)}{dz} \right)\nu(M).
\end{equation}
Here $g(\nu)$ denotes the ratio of the PDF to a Gaussian PDF:
\begin{equation}
g(\nu)=\frac{P(\nu)}{P_G(\nu)},
\end{equation}
where $P_G$ is a Gaussian PDF. Next we calculate the bias of clusters 
at the present time. The linear bias 
$\delta_{cl} =b \delta$ in Eulerian space is given by
\begin{equation}
b(M)=1- \frac{d \log n(M)}{d \delta_c}.
\end{equation}
The difference from a Gaussinan PDF is given by
\begin{equation}
b_G(M)-b(M)= \left[\frac{d \log g(\nu)}{d \nu}
\right]_{\nu(M)}\frac{1}{\sigma(M)},
\end{equation}
where the bias for a Gaussian PDF is 
\begin{equation}
b_G(M)=1+ \frac{\nu(M)^2-1}{\delta_c}
\end{equation}
(Cole $\&$ Kaiser 1989; Mo $\&$ White 1996).

If a PDF has a broad tail compared with a Gaussian PDF, therefore,
the number density of the objects $M$ evolves slowly with $z$  
and the bias of such objects is small compared with those for a Gussian PDF.
This is natural because these rare objects can be formed easily for 
a PDF with a broad tail.

\section{The $\chi^2$  model}

In this section, we investigate the $\chi_m^2$ model
( Coles and Barrow 1987, 
Moscardini et.al 1991, Weinberg and Coles 1992). In this model,
fluctuations are drawn from $\chi_m^2$ distributions.
Peebles gives one realization of this model in the framework
of inflationary universe(Peebles 1997,1999a).
In his model, CDM fields are squeezed massive scalar fields 
which have $O(N)$ symmetry. The isocurvature perturbations are 
generated during inflation. 
The density and the over-density of CDM fields
are given by
\begin{equation}
 \rho_{CDM}= \frac{\mu}{2} \sum^m_{i=1} \phi_i^2
  ,\quad \delta= \frac{1}{m \langle \phi^2 \rangle} \sum^m_{i=1} \phi_i^2 
  -1,
\end{equation}
where $m$ is the number of the CDM fields $\phi$, 
which have $O(m)$ symmetry and a Gaussian PDF and 
$\mu$ is the mass of the CDM fields.
Then, the PDF of the overdensity is given by
\begin{equation}
P(\nu)d \nu=\frac{(1+\sqrt{\frac{2}{m}} \nu)^{m/2-1}}{(\frac{2}{m})
^{(m-1)/2}\Gamma(\frac{m}{2})} \exp \left(-\frac{m}{2}\left
(1+\sqrt{\frac{2}{m}}\nu \right) \right) d \nu.
\end{equation}
If the autocorrelation function of the Gaussian field $\phi(\x)$ is
$ \xi_{\phi}=\langle \phi(\x_1) \phi(\x_2) \rangle \propto x_{12}^{-\epsilon}
(x_{12}=\vert \x_1-\x_2 \vert$), the autocorrelation function and
the reduced moments of $\delta$ are given by 
\begin{eqnarray}
\xi(x_{12})&=&\langle \delta(\x_1) \delta(\x_2) \rangle=
\left( \frac{2}{m} \right) \frac{\xi_{\phi}^2 (x_{12})}
{\langle \phi^2 \rangle}
 \nonumber\\
\lambda_N &=&(N-1)! \left( \frac{2}{m} \right)^{N/2-1} 
\frac{ (\langle x_{12}x_{23} \cdot \cdot 
x_{N-1 N} x_{N1})^{-\epsilon} \rangle}{\langle x_{12}^{-2 \epsilon} 
\rangle^{(N/2)}},
\nonumber\\
\end{eqnarray}
where angular brackets denote averaging on some smoothing length
(Peebles 1999b). 

Note that two problems seem to exist in this model. 
First, because the autocorrelation function is positive for all separations, 
the integral of $\xi(r)$ over all space does not vanish (White 1998). 
Next it can be shown that the reduced 
moments remain approximately same for a variety of smoothing scales
if $\epsilon$ is scale-invariant(Peebles 1999b, White 1998). 
This seems to be in contradiction to the central limit theorem that states
all distributions must become Gaussian when smoothed on sufficiently 
large scales. If desired, however, 
it is possible to modify the Peebles model
so that these difficulties disappear. For example, an off-centered $\chi^2_m$
model, where $\rho=\mu/2 \sum (\phi_i-\phi_{0})^2$, produces a
distribution that is non-Gaussian on small scales and Gaussian 
on large scales. This scale is determined by the mass scale 
$\phi_{0}$.
Here, we are interested in the scale relevant to
cluster formation where $\xi$ has a power-low spectrum for which $\xi>0$
and the smoothing length is not so large, so we believe that these
technical difficulties do not affect us. Indeed, our method can be 
applied to any non-Gaussian model on this scale.

In the $\chi_m^2$ model the non-Gaussianiy is represented by one parameter
$m$. As the effect of the non-Gaussianity 
depends on the broadness of the PDF, we introduce another parameter $G$ to
represent the amount of deviation from Gaussianity: 
\begin{equation}
G= \sqrt{2 \pi}
\int_3^{\infty} P(\nu) d \nu / \int_3^{\infty} e^{-\nu^2/2} d \nu,
\end{equation}
which is the likelyhood relative to Gaussian of $3 \sigma$
or rare events (Robinson et al. 1998).
$G$ is related to $m$ as follows:
\begin{center}
 \begin{tabular}{|c|c|c|c|c|c|}
 \hline 
  $m$ & 1 & 8 & 32 & 200 &$\infty$ \\
 \hline  
  $G$ & 16.3 & 7.66  & 4.02 & 2.04 & 1 \\
 \hline 
\end{tabular}
\end{center} 

\section{Evolution of abundance and correlation length of clusters}
We examine the evolution of the number density of rich clusters from 
$z=0$ to $z=1$. We consider the clusters with the mass 
$M(<R_{com} =1.5 h^{-1} \mbox{Mpc})> 8 \times 10^{14} M_{\odot}$,
where $M(<R_{com})$ represents the mass contained within 
the comoving length $R_{com}$ (Bahcall $\&$ Fan 1998).
The number density of such clusters is given by
\[
N(M(<R_{com} =1.5 h^{-1} \mbox{Mpc})> 8 \times 10^{14} M_{\odot})
=\int^{\infty}_{M_{lim}} n(M) dM.
\]
We relate the mass threshold 
$M(<R_{com} =1.5 h^{-1} \mbox{Mpc})= 8 \times 10^{14} M_{\odot}$
to the virial mass $M_{lim}$ using the density profile 
$M(<R)\propto R^{0.64}$ (Bahcall $\&$ Fan 1998). 
The result is shown for different $\Omega_m$ ($G=1$,
$\Lambda=1-\Omega_m$) 
and $G$ ($\Omega_m=1$) in Fig.1. 
The Gaussian model in the universe of $\Omega_m=1$ is strongly disfavored 
(Calberg et al 1997; Bahcall $\&$ Fan 1998).
The non-Gaussianity of fluctuations makes the evolution of the abundance 
slow. 

\begin{figure}
 \epsfysize=11.7cm  
 \begin{center}
 \epsfbox{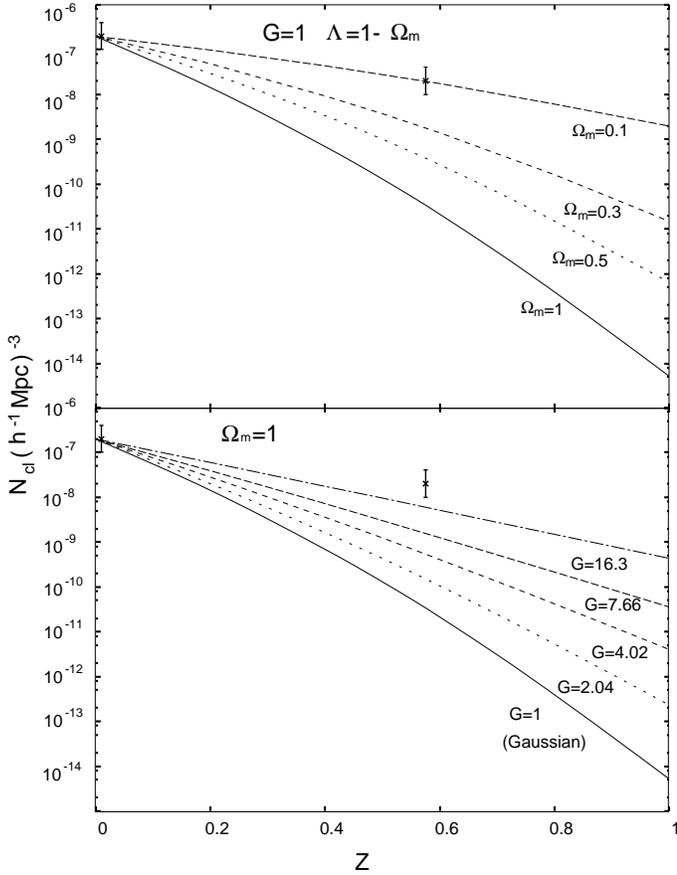}
 \end{center}
 \caption{The evolution of cluster abundance of 
$M(<R_{com} =1.5 h^{-1} Mpc)> 8 \times 10^{14} M_{\odot}$ clusters 
for a different $\Omega_m$ ($G=1$, $\Lambda=1-\Omega_m$, upper panel) and
for a different $G$ ($\Omega_m=1$, lower panel).}
\end{figure}

The correlation length $r_0$ of clusters with the mean separation 
$d=N(>R)^{-1/3}$ is obtained as follows (Robinson et al. 1998).  
Using the mass function 
we obtain the scale $R$ of clusters with the mean separation $d=N(>R)^{-1/3}$.
Then, the amplitude of two-point correlation function
of matter fluctuations at correlation length is given by
$\xi_m=1/b (R)^2$. If we know $\sigma_8$ and the spectrum index, $n$, 
we can obtain the correlation length. The bias is given by
\begin{equation}
b=1+\frac{\nu^2-1}{\delta_c(1+\sqrt{\frac{2}{m}} \nu)}.
\end{equation}
For large $G$ (small $m$), the bias becomes small 
and the correlation length becomes short.  We fix $\sigma_8=0.5$, 
$n=-1.4$ and $\Omega_m=1$; the bias and correlation length of 
the clusters with mean separation $d=30 h^{-1}$ Mpc are then given as follows:

\begin{center}
 \begin{tabular}{|c|c|c|c|c|c|}
 \hline 
 $G$ & 16.3 & 7.66  & 4.02 & 2.04 & 1 \\
 \hline  
  $b_1$ & 1.8 & 2.4  & 2.7 & 3.1 & 3.4 \\
 \hline 
  $r_0$ & 6.9 & 10.0  & 11.9 & 13.6 & 15.3 \\
 \hline 
\end{tabular}
\end{center}

\section{Constraints on $\Omega_m$, $G$ and $\sigma_8$}

We can give constraints on $\Omega_m$, $G$ and $\sigma_8$ by using 
the abundance of the rich clusters at $z=0$, $z=0.575$ and
the correlation length. We use the following data sets.
\begin{itemize}

\item The abundance of $M(<R_{com}=1.5)>8 \times 10^{14}h^{-1}
M_{\odot}$ clusters at z=0 is taken from temperature function
of $T>6.4$ kev (Bahcall $\&$ Fan 1998).
We use $2^{+2}_{-1} \times 10^{-7} h^3$Mpc$^{-3}$ (Henry $\&$ Arnaud 1991).

\item The abundance at z=0.575 is taken from the data of the Einstein 
Medium Sensitivity Survey (EMSS).
We use $n(z=0.5-0.65) =3^{+2.4}_{-1.9} \times 10^{-8} h^3$ Mpc$^{-3}$
for $\Omega_m=1$. The error bars represent $2 \sigma$ 
statistical errors (Bahcall $\&$ Fan 1998).
(For $\Omega_m<1$ we use  
$n(z=0.5-0.65) =2^{+2}_{-1} \times 10^{-8} h^3 $ Mpc$^{-3}$.)

\item The correlation length is taken from the
Automatic Plate Measuring (APM) Survey.
We use $r_0=14.2^{+0.8}_{-1.0} h^{-1}$Mpc for the clusters of mean
separation $d=30 h^{-1}$Mpc. The error bars represent $2 \sigma$
statistical errors (Croft et al. 1997).

\end{itemize}
Henceforth, for simplicity, 
we only investigate the model with a cosmological constant
and use the spectrum parameterized 
by CDM shape parameter $\Gamma$ (Bardeen et al. 1986).
We will assume $\Gamma=0.23$ (Vianna $\&$ Liddle 1998).

From the above data sets, we can give constraints on $\Omega_m$, $G$ 
and $\sigma_8$ (Fig.2).
First, consider the abundance at $z=0$. For large $G$, the clusters are
formed easily so we need small $\sigma_8$. Since $\sigma_{8 eff}=
\sigma_8 \Omega_m^{(n+3)/6}$ determines the abundance of clusters, we
need large $\sigma_8$ for small $\Omega_m$. The fit contour runs from the 
upper left (large $G$ and small $\sigma_8$) to the 
lower right(small $G$ and large $\sigma_8$) in the $\sigma_8$ vs $G$
plane and this moves to right as
$\Omega_m$ becomes small. Next, consider the abundance at $z \sim
0.6$. For $\Omega_m \sim 1$, if fluctuations are Gaussian ($G=1$), 
the abundance at $z \sim 0.6$ is too small compared with the data of
the EMSS. 
For large $G$, because of the effect of the non-Gaussianity, the abundance 
becomes consistent with the observation (Fig.1).
The fit contour runs in the plane in the same direction as
of $z=0$, but the slope is smaller. As $\Omega_m$ decreases, 
the abundance at $z \sim 0.6$ can be explained
even if fluctuations are Gaussian. Then, as $\Omega_m$ decreases, 
the fit contour moves to right more slowly than that of $z=0$. 
The region that can support these two data sets 
at the same time therefore exists in the upper left of the plane for
large $\Omega_m$ and stretches to the lower right as $\Omega_m$
becomes small. 

Next consider the correlation length. At the scale of the
correlation length, the variance of matter fluctuations is given by 
$\sigma=1/b$. For large $G$, therefore, because the bias $b$ is small, we need
large $\sigma_8$.  The fit contour runs from the lower left to the
upper right of the plane, i.e. orthogonal to the fit contours of the
abundance (Robinson et al. 1998). 
The correlation length is mainly determined by the
statistics of fluctuations, which does not change as $\Omega_m$
changes. The region that supports the correlation length and abundance at 
$z=0$ therefore moves from lower left to upper right as $\Omega_m$ decreases. 
Now, we combine all the data. For $\Omega_m \sim 1$ we cannot find a region
that will support all the data sets. As $\Omega_m$ decreases, 
we come to some $\Omega_m$, for which the region that supports all 
the data sets exists. This region moves to the upper right 
as $\Omega_m$ becomes small; \
we can then obtain the upper bound for $\Omega_m$ and the lower 
bound for $\sigma_8$ and $G$.

From the above data sets, the upper bound $\Omega_m<0.5$ and the lower bound
$G >2$ and $\sigma_8 >0.7$ are obtained. 
Particularly in the universe of $\Omega_m=0.3$, 
which is favored by other observations, 
non-Gaussianity of the order $G \sim 4$ is favored.
\begin{figure*}
 \epsfysize=12.4cm  
 \begin{center}
 \epsfbox{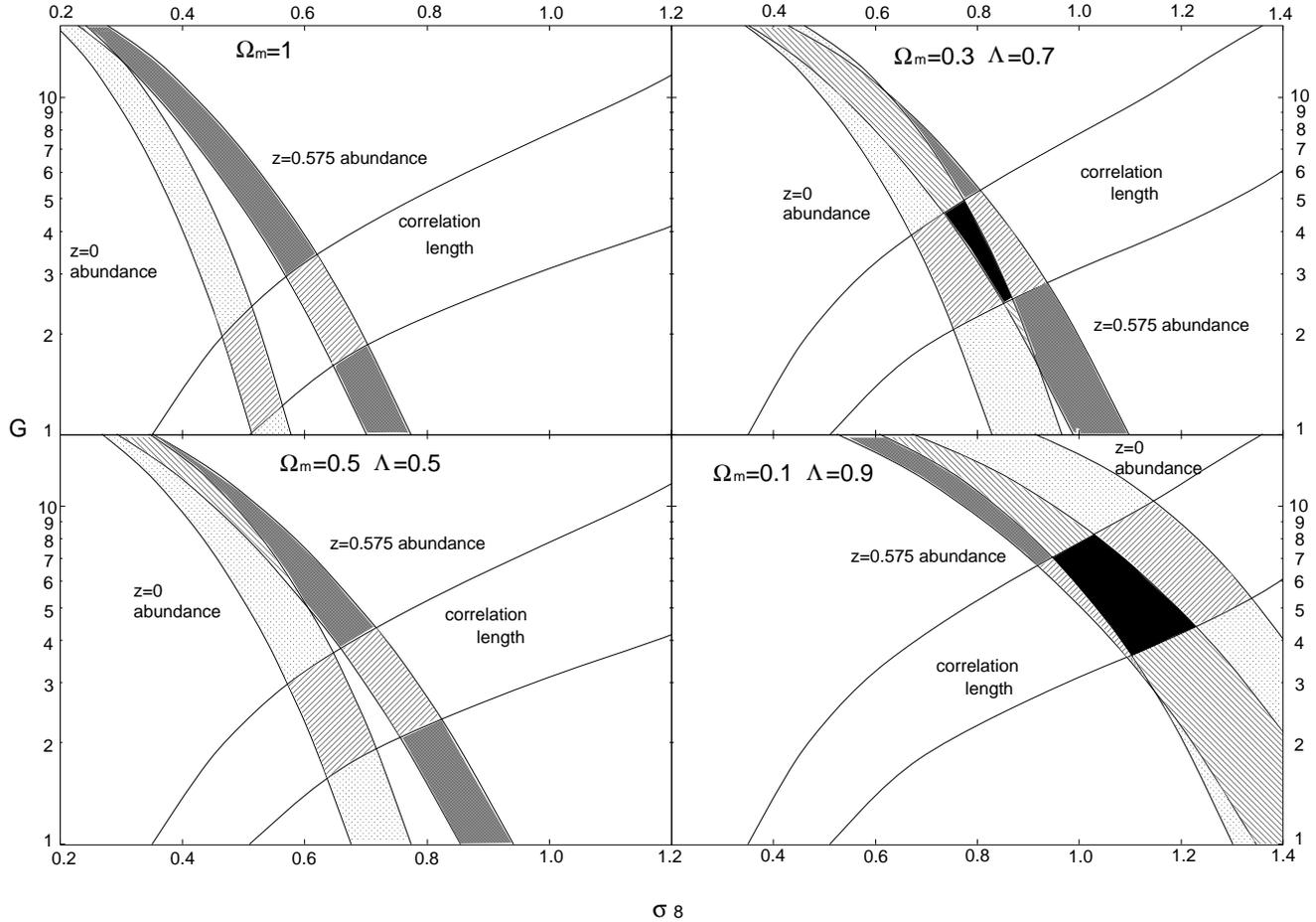}
 \end{center}
 \caption{Fit contours of the three data sets in $\sigma_8$ vs $G$ plane.} 
 \end{figure*}

\begin{figure}
 \epsfysize=7cm  
 \begin{center}
 \epsfbox{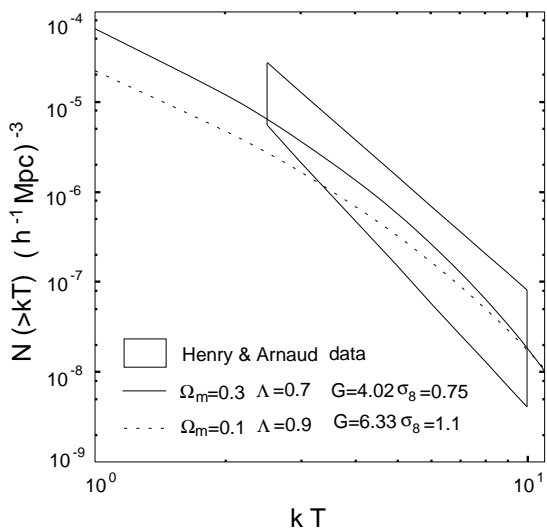}
 \end{center}
\caption{The cumulative cluster temperature function for the 
$\chi^2_m$ model at the present time.}
 \end{figure}

The shape of the cluster temperature function could possibly 
give another constraint on the model (Kitayama and Suto 1996).
We show the temperature function at the present time
for the parameters that are consistent with all three data sets
($\Omega_m=0.3, \Lambda=0.7, G=4.02, \sigma_8=0.75$ 
and $\Omega_m=0.1, \Lambda=0.9, G=6.33, \sigma_8=1.1$ (Fig.3)).   
Here we convert the virial mass $M_{v}$ to X-ray temperature using 
\[
kT = \frac{7.75}{\beta} \left( \frac{M_{v}}{10^{15}h^{-1}M_{\odot}}
\right)^{2/3}(1+z)
\left( \frac{\Omega_m}{ \Omega(z) } \right)^{1/3}
\left (\frac{\triangle_v}{178} \right)^{1/3} \mbox{kev},
\]
where $\triangle_v$ is the virial over-density  
($\triangle_v=\rho_{virial}/\rho_c$) and we take $\beta=1$ (Eke et al. 1996;
1998; Borgani et al 1998). The shape of the temperature function agrees well
with the data of Henry and Arnaud for high temperatures 
(Henry \& Arnaud 1991, Henry 1997). Note that as $\Omega_m$ decreases, 
the non-Gaussianity and $\sigma_8$ becomes large, so the
slope of the temperature function becomes smaller. Then, the difference 
in the abundance for different $\Omega_m$ is larger for low temperatures.
This provides us with the possibility of obtaining another constraint 
on the low-density universe model.

\section{Conclusion}
We have given constraints on the $\chi_m^2$ model, where
fluctuations are drawn from $\chi_m^2$ distributions.
We extended the Press-Scheter theory to calculate the abundance and 
bias of clusters. The non-Gaussianity of fluctuations makes the
evolution of the abundance slow. 
The model with non-Gaussian fluctuations of 
the order $G\sim 16$ $(m=1)$ in $\Omega_m=1$ 
universe has roughly the same evolution of
the number density as a Gaussian model in a $\Omega_m=0.2$ universe, 
where $G$ represents the non-Gaussianity of the fluctuations
($G=1$ for Gaussian). 
On the other hand, the strong non-Gaussianity of fluctuations 
make the correlation length too short. Combining the
three data sets, the abundance of the clusters at $z=0$ and  
$z \sim 0.6$ and the correlation length, 
constraints on the non-Gaussianity and the
amplitude of fluctuations and the density parameter have been obtained.
We have shown that the upper bound for $\Omega_m$ and the lower bound for the
non-Gaussianity and the amplitude of the fluctuations can be given.
For the spectrum parameterized by CDM shape parameter $\Gamma=0.23$, 
we have obtained an upper bound for the density parameter $\Omega_m \sim 0.5$ 
and lower bounds for the amplitude $\sigma_8 \sim 0.7$ 
and the non-Gaussianity $G \sim 2$ $(m \sim 200)$ of fluctuations. 
In the universe with $\Omega_m=0.3$ and $\Lambda=0.7$, 
non-Gaussianity of the order $G \sim 4$ $(m \sim 30)$ is preferred.

\section{acknowledgment}
The authors are grateful to J.Robinson for useful comments.
The work of J.S. was supported by Monbusho Grant-in-Aid No.10740118
and the work of K.K. was supported by JSPS Research Fellowships for
Young Scientist No.04687.

\appendix
\section{Cross-correlation between scales}
In this appendix, we show that if $R_0 \gg R_1$, the relation
\begin{equation}
\langle \delta_1^p \delta_0^{n-p} \rangle = \langle \delta_0^n \rangle.
\end{equation}
is satisfied.
This is the extension of the relation for Gaussian initial conditions
which was derived by Bower under the assumption that the volume of the 
region $0$ is sufficiently larger than that of the region $1$ (Bower 1991). 
The smoothed overdensity in the region of scale $R$ is
\begin{equation}
  \delta(\x,R)=\int\frac{d^3\k}{(2\pi)^3} \delta_{\mbox{\boldmath$k$} }
W_R(k R)e^{-i \k \cdot \x},
\end{equation}
where $W_R(kR)$ is a smoothing function. The n-point functions of the
random field $\delta_{\k}$ are expressed as
\begin{eqnarray}
\langle \delta_{\k_1} \delta_{\k_2} \rangle &=& \xi_2(\k_1,\k_2) ,\nonumber\\
\langle \delta_{\k_1} \delta_{\k_2} \delta_{\k_3} \rangle 
&=& \xi_3(\k_1,\k_2,\k_3) \nonumber\\ 
\langle \delta_{\k_1} \delta_{\k_2} \delta_{\k_3} \delta_{\k_4}\rangle 
&=& \xi_4(\k_1,\k_2,\k_3,\k_4) + \xi_2(\k_1,\k_2)\xi_2(\k_3,\k_4)
\nonumber\\
&& + cycl. \: (\mbox{2-terms}).
\end{eqnarray}
Consider the region $1$ with scale $R_1$ centered $\x_1$ contained 
in the region $0$ with scale $R_0$ centered $\x_0$.
The variance of the overdensity in the region $i$ is
\begin{equation}
\sigma^2(R_i) = \langle \delta_i^2 \rangle
=\int\frac{d^3\k}{(2\pi)^3} W_R^2(kR_i)P(k), 
\end{equation}
where $P(k)$ is defined by $\xi(\k_i,\k_j)=\delta_D(\k_i+\k_j)P(k_i)$.
The covariance is written as
\begin{eqnarray}
\lefteqn{\sigma^2(R_0 R_1)= \langle \delta(R_0) \delta(R_1) \rangle}
\nonumber\\
&& =\int \frac{d^3\k}{(2\pi)^3}
P(k) W(kR_0)W(kR_1) \langle e^{-i \:\k \cdot \y} \rangle_{1 \in 0},
\end{eqnarray}
where $\y=\x_0-\x_1$.
Averaging over $\y$ is done as follows: select a random set of points $\y$ 
for which all points in the region $1$ are independent 
and enclosed in the region $0$, then average over this random set of $\y$.
The averaged term can be written to appear like another window function
\begin{eqnarray}
\tilde{W}_{\tilde{R}}(k \tilde{R}) &=&
\langle e^{-i \k \cdot \y} \rangle_{1 \in 0} \nonumber\\
&=& \int_{\tilde{R}} d^3 \y \:w(\y) e^{-i \k \cdot \y} / 
\int_{\tilde{R}} d^3 \: \y w(\y), 
\end{eqnarray}
where $\tilde{R}$ is the scale covered by the points $\y$ for which 
region $1$ lies in region $0$. The weight function $w(\y)$ is defined
so that every point in the region $0$ has an equal chance of contributing
to the average density measured in the region 1. Then we can write 
the real-space window function of the region $0$ as the convolution
\begin{equation}
W_{R_0}(\x)=\int d^3 \x' W_{R_1}(\x) \tilde{W}_{\tilde{R}}(\x-\x').
\end{equation}
In Fourier space
\begin{equation}
W_{R_0}(kR_0)=\tilde{W}_{\tilde{R}}(k \tilde{R}) W_{R_1} (kR_1).
\end{equation}
Note that it is possible to find real weighting function only if 
the region $1$ is sufficiently small compared to the region $0$, i.e.
$R_0 \gg R_1$.
If this is the case, the covariance is given by
\begin{equation}
\sigma^2(R_0 R_1) =\int\frac{d^3\k}{(2\pi)^3} W_R^2(kR_i)P(k)=\sigma^2(R_0).
\end{equation}
Extensions to the higher moments are quite similar. For example, 
\begin{eqnarray}
\lefteqn{\langle \delta_1^p \delta_0^{n-p} \rangle} \nonumber\\
&=&\int d^3 \k_1 \cdot \cdot d^3 \k_n \delta_D^3(\k_1+\cdot \cdot \k_n)
\langle \delta_{\k_1} \cdot \cdot \cdot \delta_{\k_n} \rangle \nonumber\\
&& \times W_{R_0}(k_1 R_0) \cdot \cdot \cdot W_{R_0}(k_{n-p} R_0)\nonumber\\
&& \times [W_{R_1}(k_{n-p+1} R_1)\tilde{W}_{\tilde{R}}(k_{n-p+1}\tilde{R})]
\cdot \cdot \cdot  \nonumber\\
&& \times 
[W_{R_1}(k_{n} R_1)\tilde{W}_{\tilde{R}}(k_{n} \tilde{R})]
\nonumber\\
&=& \int d^3 \k_1 \cdot \cdot  d^3 \k_n \delta_D^3(\k_1+ \cdot \cdot \k_n)
\langle \delta_{\k_1} \cdot \cdot \cdot \delta_{\k_n} \rangle \nonumber\\
&& \times W_{R_0}(k_1 R_0) \cdot \cdot \cdot W_{R_0}(k_n R_0)
\nonumber\\
&=&\langle \delta_0^n \rangle.
\end{eqnarray}

\section{Conditional probability}
We will derive the Edgeworth expansion 
(Bernardeau $\&$ Kofman 1995; Juszkiewicz et al. 1995)
for the conditional probability of $\delta_0$ and
$\delta_1$ in the case that $R_0 \gg R_1$ is satisfied. 
The generating function for joint moments
\begin{equation}
C(\mu,\nu)= \int d \delta d \theta e^{i \mu \delta+i \nu \theta}
P(\delta, \theta)
\end{equation}
gives rise to
\begin{equation}
\langle \delta^n \theta^m \rangle =\left. \frac{\partial^{n+m} C(\mu, \nu)}
{\partial(i \mu)^n \partial (i \nu)^m} \right|_{\mu=\nu=0}.
\end{equation}
The characteristic function becomes
\begin{equation}
K(\mu,\nu)= \log C(\mu,\nu)=\sum^{\infty}_{(n,m) \neq (0,0)}
\frac{(i \mu)^n (i \nu)^m }{n! m!} \lambda_{nm} ,
\end{equation}
where connected joint moments
\begin{equation}
\lambda_{mn}= \langle \delta^n \theta^m \rangle_c
\end{equation}
are used. The inverse transformation gives the conditional probability
function
\begin{eqnarray}
\lefteqn
{P(\delta,\theta)=\frac{1}{(2 \pi)^2} \int d \mu d\nu e^{-i\mu \delta
-i \nu \theta} e^{K(\mu,\nu)}} \nonumber\\
&&=\frac{1}{(2 \pi)^2} \int  d \mu d\nu e^{-i\mu \delta
-i \nu \theta -\frac{1}{2}\mu^2-\frac{1}{2} \nu^2-r \mu \nu} 
e^{\Sigma(\mu,\nu)} ,
\end{eqnarray}
where
\begin{equation}
\Sigma(\mu,\nu)=\sum^{\infty}_{N=3} \sum_{m+n=N} \frac{(i \mu)^n(i
  \nu)^m}{n! m!} \lambda_{nm}
\end{equation}
is the nonlinear part of the characteristic function and $r=\langle
\delta \theta \rangle$ is a cross-correlation. Here we rescale 
$\delta$ and $\theta$ so as to have a unit variance. Simple calculation
yields
\begin{equation}
P(\delta, \theta)=e^{\Sigma(i \frac{\partial}{\partial \delta},
i \frac{\partial}{\partial \theta})} \:\frac{1}{2 \pi \sqrt{1-r^2}}\:
e^{-\frac{1}{2} \frac{(\delta-r \theta)^2}{1-r^2}-\frac{1}{2}
  \theta^2}.
\end{equation}
To proceed further, we use the relation derived in appendix A:
\begin{equation}
\langle \delta^N \rangle=\lambda_{N0}, \langle \delta^p \theta^{N-p}
\rangle=r^p \lambda_{0N},\langle \theta^N \rangle= \lambda_{0N}.
\end{equation}
Now, let us define
\begin{equation}
\alpha=\frac{\delta-r \theta}{\sqrt{1-r^2}} ,\quad \beta=\theta.
\end{equation}
We then obtain
\begin{eqnarray}
P(\delta,\theta)&=& \exp \left(\sum^{\infty}_{N=3} 
\left[ \frac{1}{N!}( \lambda_{N0}-\lambda_{0N} r^N) 
\left(-\frac{1}{\sqrt{1-r^2}} \frac{\partial}{\partial \alpha}
\right)^N  \right]\right)  \nonumber\\
&& \times \frac{1}{\sqrt{2 \pi}} e^{-\frac{1}{2} \alpha^2} P(\theta),
\end{eqnarray}
where 
\begin{eqnarray}
P(\theta)=\exp{ \left(\sum^{\infty}_{N=3} 
\left[ \frac{\lambda_{0N}}{N!} 
\left(-\frac{1}{\partial \beta}
\right)^N  \right] \right)
\frac{1}{\sqrt{2 \pi}}} e^{-\frac{1}{2} \beta^2} \nonumber\\
=\frac{1}{\sqrt{2 \pi}} e^{-\frac{\theta^2}{2}} 
\left( 1+ \frac{\lambda_3}{3!} H_3(\theta)+O(\lambda_3^2, \lambda_4) \right).
\end{eqnarray}
Here we use the definition of the Hermite functions 
\begin{equation}
H_n(x)=e^{\frac{x^2}{2}} \left( - \frac{d}{dx} \right)^n
  e^{-\frac{x^2}{2}}.
\end{equation}
Then the conditional probability function is given by
\begin{eqnarray}
\lefteqn{P(\delta |\theta) =\frac{P(\delta,\theta)}{P(\theta)}} \nonumber\\
&& 
=\exp \left( \sum^{\infty}_{N=3} 
\left[ \frac{1}{N!}( \lambda_{N0}-\lambda_{0N} r^N) 
\left(-\frac{1}{\sqrt{1-r^2}} \frac{\partial}{\partial \alpha}
\right)^N  \right] \right)  \nonumber\\
&& \times \frac{1}{\sqrt{2 \pi}} e^{-\frac{1}{2} \alpha^2}.
\end{eqnarray}
Consider the small region $1$ contained in a large region $0$. 
Defining the variance $\sigma_1$ for $\delta_1$ and $\sigma_0$ for
$\delta_0$ and rescaling yields 
the Edgeworth formula for conditional probability
\begin{eqnarray}
P(\delta_1 |\delta_0)= \frac{1}{\sqrt{2 \pi}
  \sqrt{\sigma_1^2-\sigma_0^2}}
\exp \left[ -\frac{(\delta_1- \delta_0)^2}{2 (\sigma_1^2 -\sigma_0^2)}
\right] \nonumber\\
\times
\left( 1+ \frac{\langle \delta_1^3 \rangle-\langle \delta_0^3
    \rangle}{3!( \sigma_1^2-\sigma_0^2)^{3/2}} H_3 \left(
\frac{\delta_1-\delta_0}{\sqrt{\sigma_1^2-\sigma_0^2}}\right) + \cdot \cdot
\cdot \right).
\end{eqnarray}
Using the relation
\begin{equation}
\langle \delta_1^n \rangle \gg \langle \delta_0^n \rangle,
\end{equation}
the conditional probability is reduced to
\begin{equation}
P(\delta_1 | \delta_0)=P(\delta_1-\delta_0).
\end{equation}

\label{lastpage}

\end{document}